\documentclass[11pt]{article}
\usepackage{mymoriond1,epsfig}

\def\be{\begin{equation}}
\def\ee{\end{equation}}
\def\bea{\begin{eqnarray}}
\def\eea{\end{eqnarray}}

\newcommand{\lsim}{
\mathrel{\hbox{\rlap{\hbox{\lower4pt\hbox{$\sim$}}}\hbox{$<$}}}}

\newcommand{\gsim}{
\mathrel{\hbox{\rlap{\hbox{\lower4pt\hbox{$\sim$}}}\hbox{$>$}}}}

\begin{document}
\begin{flushright}
\begin{tabular}{l}
IPPP/07/07\\
DCPT/07/14
\end{tabular}
\end{flushright}

\vspace*{4cm}

\title{\boldmath PROBING NEW PHYSICS THROUGH $B_s$ MIXING}

\author{PATRICIA BALL}

\address{IPPP, Department of Physics, University of Durham,\\
Durham DH1 3LE, England}

\maketitle\abstracts{
I discuss the interpretation of the recent experimental
data on $B_s$ mixing in terms of model-independent new-physics
parameters.\\[2cm]
{\it Invited Talk given at XLIInd Rencontres de Moriond,
  Electroweak Interactions and Unified Theories, La Thuile, Italy,
  March 2007}}

\newpage

\section{Introduction}

One of the most promising ways to detect the effects of new physics (NP) on 
$B$ decays is to look for deviations of flavour-changing neutral-current 
(FCNC) processes from their Standard Model (SM) predictions;
FCNC processes only occur at the loop-level in the SM and
hence are particularly sensitive to NP virtual particles and
interactions. A prominent example that has received extensive
experimental and theoretical attention is $B^0_q$--$\bar B^0_q$ mixing
($q\in\{d,s\}$), which, in the SM, is due to box diagrams with $W$-boson
and up-type quark exchange. In the language of effective field theory, 
these diagrams
induce an effective local Hamiltonian, which causes $B^0_q$ and $\bar B^0_q$ 
mesons to mix and generates a $\Delta B=2$ transition:
\begin{equation}\label{eq1}
\langle B_q^0| {\cal H}^{\Delta B=2}_{\rm eff} | \bar B_q^0\rangle = 2 M_{B_q}
M_{12}^q\,,
\end{equation}
where $M_{B_q}$ is the $B_q$-meson mass. Thanks to $B^0_q$--$\bar B^0_q$
mixing, an initially present $B^0_q$ state evolves into a time-dependent linear
combination of $B^0_q$ and $\bar B^0_q$ flavour states. The
oscillation frequency
of this phenomenon is characterized by the mass difference of the ``heavy"
and ``light" mass eigenstates,
\begin{equation}\label{DM-def}
\Delta M_q\equiv M_{\rm H}^{q}-M_{\rm L}^{q} = 2 |M_{12}^q|\,,
\end{equation}
and the CP-violating mixing phase 
\begin{equation}\label{phiq-def}
\phi_q = \arg M_{12}^q\,,
\end{equation}
which enters mixing-induced CP violation. While the mass difference in
the $B_d$ system has been known for a long time, $\Delta M_s$ has only
been measured in 2006, by the CDF collaboration, with the result
\cite{CDFms}
\begin{equation}\label{4}
\Delta M_s = [17.77\pm 0.10({\rm stat})\pm 0.07({\rm syst})]{\rm
  ps}^{-1}.
\end{equation}
In Ref.~\cite{PB4}, we have discussed the impact of this result on
a model-independent parametrisation of NP in the $B_q$
system. In the meantime, experimental information has become available
also for the mixing phase in the $B_s$ system
\cite{D01,D02,D03,D0talk}. In these proceedings, we update the
constraints obtained on NP in the $B_s$ system by including
this additional information.

In the SM, $M_{12}^q$ is given by
\begin{equation}\label{5}
M_{12}^{q,{\rm SM}} = \frac{G_{\rm F}^2M_W^2}{12\pi^2}M_{B_q}\hat{\eta}^{B}
\hat B_{B_q}f_{B_q}^2(V_{tq}^\ast V_{tb})^2 S_0(x_t)\,,
\end{equation}
where $G_{\rm F}$ is Fermi's constant, $M_W$ the mass of the $W$ boson, 
$\hat{\eta}^{B}=0.551$ a short-distance QCD correction (which is the same for
the $B^0_d$ and $B^0_s$ systems) \cite{jamin}, whereas
the bag parameter
$\hat B_{B_q}$ and the decay constant $f_{B_q}$ are non-perturbative 
quantities.
$V_{tq}$ and $V_{tb}$ are elements of the 
Cabibbo--Kobayashi--Maskawa (CKM) matrix, and 
$S_0(x_t\equiv \overline{m}_t^2/M_W^2)=2.32\pm0.04$ with $\overline{m}_t(m_t) =
(163.4\pm 1.7)\,{\rm GeV}$, Ref.~\cite{top}, describes
the $t$-quark mass dependence of the box diagram with internal 
$t$-quark exchange; the contributions of internal $c$ and $u$ quarks
are suppressed by  $(m_{u,c}/M_W)^2$, by virtue of the GIM
mechanism. Thanks to the suppression of light-quark loops, $M_{12}^q$ is
dominated by short-distance processes and sensitive to NP.

In the SM, the mixing phase in the $B_s$ system is given by
\begin{equation}
\phi_s^{\rm SM} = -2 \lambda^2 R_b\sin\gamma \approx -2^\circ\,,
\end{equation}
where $\gamma$ is one of the angles of the unitarity
triangle (UT), $\lambda$ is the Wolfenstein parameter and 
\begin{equation}
R_b = \left( 1 - \frac{\lambda^2}{2}\right)
\frac{1}{\lambda}\,\left| \frac{V_{ub}}{V_{cb}}\right|.
\end{equation}
Up-to-date values of $\gamma$ from various sources can be found in
Ref.~\cite{gamma}, whereas $|V_{ub}|$ and $|V_{cb}|$ can be in found
Refs.~\cite{vub} and \cite{vcb}, respectively. The corresponding
results from global fits can be found in Ref.~\cite{global}.

\begin{figure}
$$\epsfxsize=0.47\textwidth\epsffile{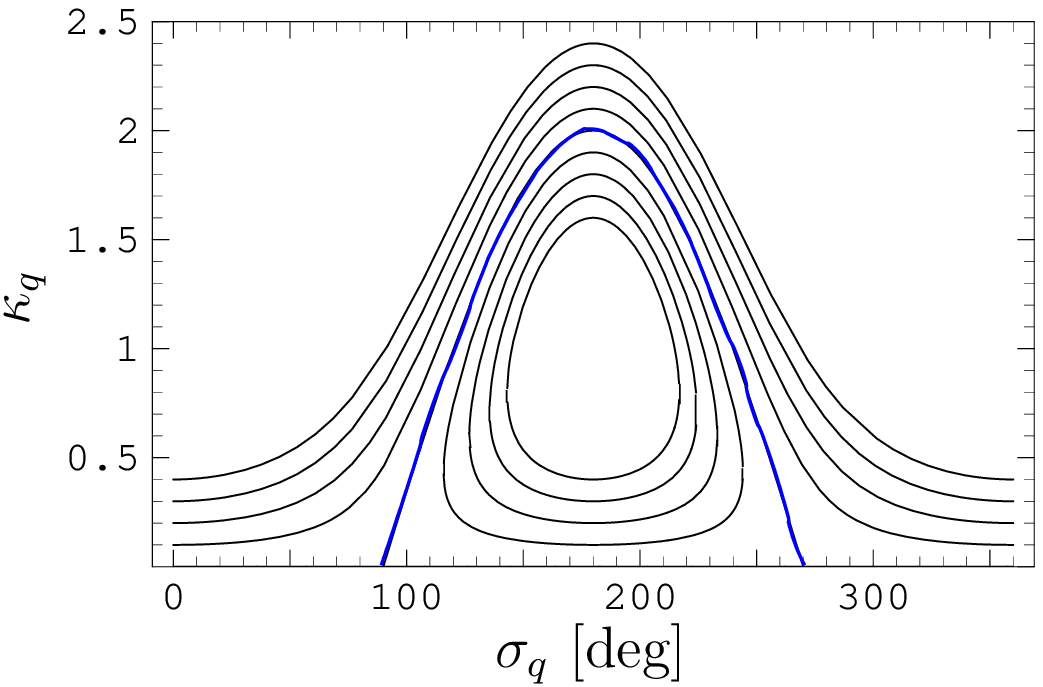}\quad
\epsfxsize=0.47\textwidth\epsffile{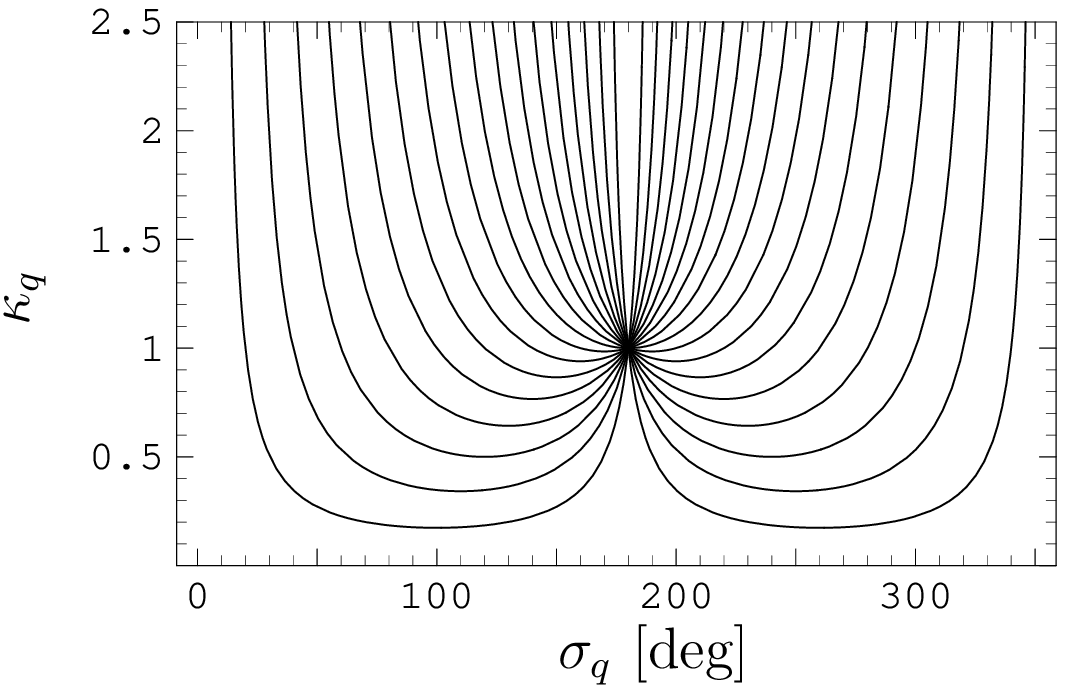}
$$
 \vspace*{-1truecm}
\caption[]{\small Lines of constant $\rho_s$ (left) and constant
  $\phi_s^{\rm NP}$ (right) in the $\sigma_s$-$\kappa_s$ plane. Blue
  line: $\rho_s\equiv 1$.}\label{fig1}
\end{figure}

In the presence of NP, the matrix element $M_{12}^q$ can be
written, in a model-independent way, as
$$M_{12}^q = M_{12}^{q,{\rm SM}} \left(1 + \kappa_q e^{i\sigma_q}\right)\,,$$
where the real parameter $\kappa_q\geq 0$ measures the ``strength'' of
the NP contribution with respect to the SM, whereas 
$\sigma_q$ is a new CP-violating
phase. 
Relating $\kappa_s$ and $\sigma_s$ to $\Delta M_s$, one has
\begin{equation}
\rho_s\equiv \left|\frac{\Delta M_s}{\Delta M_s^{\rm SM}}\right| =
\sqrt{1+ 2 \kappa_s\cos \sigma_s + \kappa_s^2}\,.
\end{equation}
The lines of $\rho_s=\,$const.\ in the $\sigma_s$-$\kappa_s$ plane are
shown in Fig.~\ref{fig1}. The blue line $\rho_s=1$ illustrates that even if
the experimental value of $\Delta M_s$ {\em  coincides} with the SM
expectation, it is wrong to conclude that there is no NP in
$B_s$ mixing -- in fact, in this case the NP amplitude can be  larger
than the SM amplitude, i.e.\ $\kappa_s>1$, if SM and NP contributions
differ by a phase $\sigma_s$ between 120$^\circ$ and 240$^\circ$.

In order to obtain $\rho_s$ from the experimental result (\ref{4}),
one has to determine $\Delta M_s^{\rm SM}$. In addition to the input
parameters listed after (\ref{5}), one also needs the CKM matrix
elements $|V_{ts}^* V_{tb}|$ and the hadronic matrix element $\hat
B_{B_s} f_{B_s}^2$. The former is accurately known in terms of
$|V_{cb}|$ and $\lambda$ and reads \cite{PB4}
\begin{equation}
|V_{ts}^* V_{tb}| = \left\{ 1 - \frac{1}{2}\,(1-2 R_b\cos\gamma)
 \lambda^2 + {\cal O}(\lambda^4)\right\} |V_{cb}^* V_{tb}| = (41.3\pm
 0.7)\times 10^{-3}\,.
\end{equation}
The hadronic matrix element $\hat B_{B_s} f_{B_s}^2$ has 
been the subject of numerous lattice calculations,
both quenched and unquenched,  using various lattice actions and  
implementations of both heavy and light quarks. The current front 
runners are unquenched
calculations with 2 and 3 dynamical quarks, respectively, and Wilson
or staggered light quarks. Despite tremendous progress in recent
years, the results still suffer from a variety of uncertainties which
is important to keep in mind when interpreting and using lattice
results. The most recent (unquenched) simulation by the JLQCD collaboration
\cite{JLQCD}, with non-relativistic $b$ quarks and two flavours of
dynamical light (Wilson) quarks, yields 
\begin{equation}\label{JLQCD}
\left.f_{B_s}\hat{B}_{B_s}^{1/2}\right|_{\rm JLQCD} = (0.245\pm
0.021^{+0.003}_{-0.002})\,{\rm GeV}\,,
\end{equation}
where the first error includes uncertainties from statistics and
various systematics, where\-as the second, asymmetric error comes
from the chiral extrapolation from unphysically large light-quark
masses to the $s$-quark mass. 

More recently, (unquenched) simulations with three dynamical flavours 
have become possible using staggered quark actions. The HPQCD
collaboration obtains \cite{HPQCD}
\begin{equation}\label{HPQCD}
\left.f_{B_s} \hat B_{B_s}^{1/2}\right|_{\rm HPQCD} = (0.281\pm
0.021)\,{\rm GeV}\,.
\end{equation}
where all errors are added in quadrature. 

Although we shall use both (\ref{JLQCD}) and (\ref{HPQCD}) in our
analysis, we would like to stress that the
errors are likely to be optimistic. There is the question of
discretisation effects (JLQCD uses data obtained at only one lattice
spacing) and the renormalisation of matrix elements (for lattice
actions without chiral symmetry, the axial vector current is not
conserved and $f_{B_q}$ needs to be renormalised), which some argue should
be done in a non-perturbative way \cite{alpha}. Simulations with
staggered quarks also face potential problems with unitarity,
locality and an odd number of flavours (see, for instance,
Ref.~\cite{sharpe}). A confirmation of the HPQCD results by
simulations using the (theoretically better understood) Wilson action
with small quark masses will certainly be highly welcome.

\begin{figure}
$$\epsfxsize=0.47\textwidth\epsffile{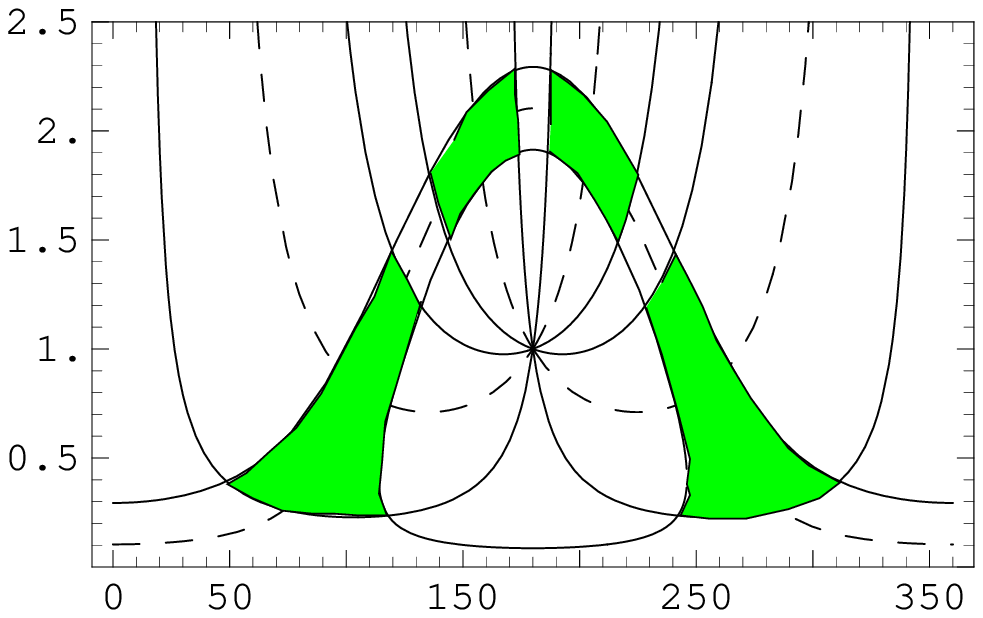}\quad
\epsfxsize=0.47\textwidth\epsffile{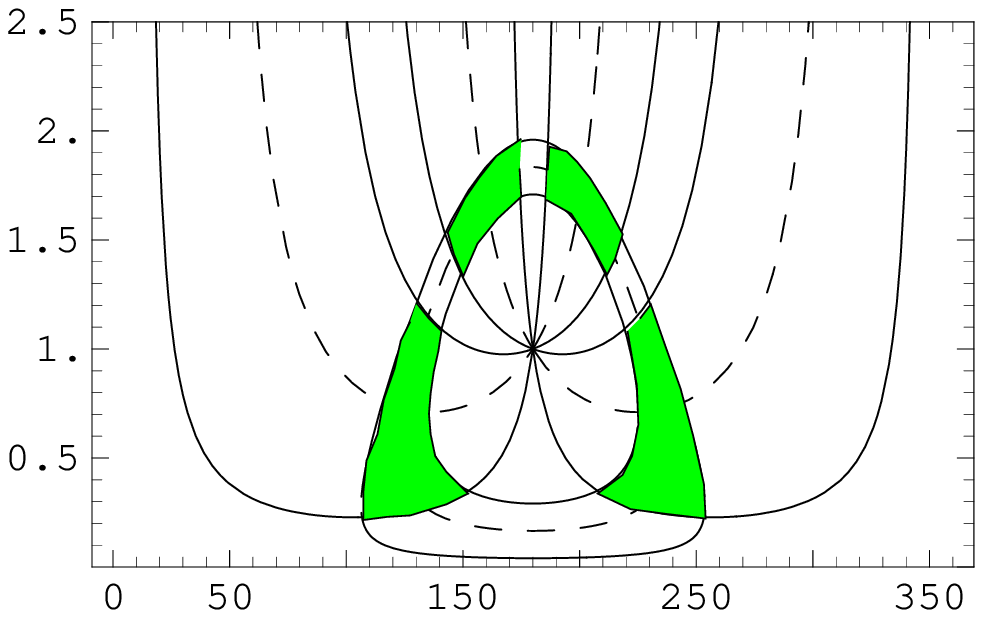}
$$
 \vspace*{-1truecm}
\caption[]{\small Allowed $1\sigma$ regions (green/grey) in the 
$\sigma_s$--$\kappa_s$ plane.
Left panel: JLQCD lattice results, Eq.~(\ref{JLQCD}). 
Right panel: HPQCD lattice 
results, Eq.~(\ref{HPQCD}). The four allowed regions correspond to the
fourfold ambiguity in the determination of $\phi_s^{\rm NP}$ from
data.}\label{fig2}
\end{figure}

With the above input parameters, one finds
\begin{equation}
\begin{array}[b]{rl@{\quad}rl}
\left. \Delta M_s \right|_{\rm JLQCD} = & (16.1\pm 2.8)\,{\rm
  ps}^{-1}, & 
\left. \Delta M_s \right|_{\rm HPQCD} = & (21.3\pm 3.2)\,{\rm
  ps}^{-1},\\[5pt]
\left.\rho_s\right|_{\rm JLQCD} = & 1.10\pm 0.19, &
\left.\rho_s\right|_{\rm HPQCD} = & 0.83\pm 0.13.
\end{array}
\end{equation}
The corresponding constraints in the $\sigma_s$-$\kappa_s$ plane are
shown in Fig.~\ref{fig2}.

In order to further constrain the NP parameter space, one needs to include
 information on the NP CP-violating phase $\phi_s^{\rm NP}$. At the
time Ref.~\cite{PB4} was written, no such information was
available. In the meantime, $\phi_s^{\rm NP}$ has been constrained
from measurements by the D0 collaboration, of the CP-asymmetry in
flavour-specific (semileptonic) $B_s$ decays \cite{D01} and 
the time-dependent angular analysis of untagged $B_s\to J/\psi\phi$
decays \cite{D02}. These measurements can be translated, using
supplementary information on the semileptonic asymmetry in $B_d$
decays, in the following results for $\Delta\Gamma_s$ and
$\phi_s$: \cite{D03,D0talk}
\begin{equation}\label{13}
\Delta\Gamma_s \equiv \Gamma_L - \Gamma_H = (0.13\pm 0.09)\,{\rm
  ps}^{-1},\quad \phi_s = -0.70^{+0.47}_{-0.39}\,.
\end{equation}
These results actually are determined only up to a 4-fold ambiguity
for $\phi_s$ and the sign of $\Delta\Gamma_s$: $\phi_s\to \pm \phi_s$
for $\Delta\Gamma_s>0$ and $\phi_s\to\pm (\pi-\phi_s)$ for
$\Delta\Gamma_s<0$. As the SM prediction for $\phi_s$ is close to 0
\footnote{Which also implies that we do not have to distinguish our
  definition of $\phi_s$ as ${\rm arg} M_{12}^s$ from the definition
  used by the D0 collaboration, $\phi_s = {\rm
  arg}(-M_{12}^s/\Gamma_{12}^s)$.}, we can identify this result with
  $\phi_s^{\rm NP}$.
The combined constraints posed by $\Delta M_s$ and
$\phi_s^{\rm NP}$ on the new-physics parameters $\kappa_s$, $\sigma_s$ are
shown as green areas in Fig.~\ref{fig2}, including the 4-fold
ambiguity. It is evident that at present the experimental error of
  $\phi_s^{\rm NP}$ is too large to considerably 
reduce the area constrained by
  $\Delta M_s$ alone. The ambiguity can be reduced to a 2-fold one if some
theory-input about the signs of $\cos\delta_{1,2}$ is used, where
$\delta_{1,2}$ are the strong phases involved in the angular analysis
of $B_s\to J/\psi \phi$, Ref.~\cite{dunietz}.
At LHCb, it will be possible to study the time-dependence of flavour-tagged 
$B_s$ decays, which gives access to the mixing-induced asymmetry and
allows one to reduce the number of discrete ambiguities without input from
theory. 

The above results can be compared with the following recent 
theory prediction for 
$\Delta\Gamma_s$, which is based on an improved operator product
expansion of $\Gamma_{12}^s$, the off-diagonal element of the $B_s$
decay matrix: \cite{LN06}
\begin{equation}
  \Delta\Gamma_s^{\rm th} = (0.096\pm 0.039)\,{\rm ps}^{-1},
\end{equation}
which agrees with the experimental result (\ref{13}) within errors.
Ref.~\cite{LN06} also
contains a detailed discussion of the theoretical predictions for
flavour-specific CP asymmetries both in the $B_d$ and $B_s$ system and
the constraints on NP in $B_s$ mixing 
extrated from all available experimental data.

Let us conclude with a few remarks concerning the prospects for the
search for NP through $B^0_s$--$\bar B^0_s$ mixing at the LHC. This
task will be very challenging if essentially no CP-violating effects
will be found in $B_s\to J/\psi \phi$ (and similar decays). 
On the other hand, even a small phase 
$\phi_s^{\rm NP}\approx-10^\circ$ would lead
to CP asymmetries at the $-20\%$ level, which could be unambiguously detected 
after a few years of data taking, and would not be affected by
hadronic uncertainties. Ref.~\cite{hunen} quotes
a sensitivity to $\phi_s$ of $\sigma(\phi_s) = 1.2^\circ$ for an
integrated luminosity of 2fb$^{-1}$ at LHCb and a sensitivity
$\sigma(\Delta \Gamma_s/\Gamma_s)\sim 0.01$ for both LHCb and
Atlas/CMS (at 30fb$^{-1}$). 
Conversely, the measurement of such an asymmetry would 
allow one to establish a lower bound on the strength of the NP
contribution -- even if hadronic uncertainties still preclude a direct 
extraction of this contribution from $\Delta M_s$ -- and to
dramatically reduce the allowed region in the NP parameter space. 
In fact, the situation may be even more promising, as specific scenarios of NP 
still allow large new phases in $B^0_s$--$\bar B^0_s$ mixing, also after the 
measurement of $\Delta M_s$, see, for instance, Refs.~\cite{emi,JMF}.

In essence, the lesson to be learnt from this discussion 
is that NP may actually be hiding in $B^0_s$--$\bar B^0_s$
mixing, but is still obscured by parameter uncertainties, some of which will 
be reduced by improved statistics at the LHC, whereas others require dedicated 
work of, in particular, lattice theorists. The smoking gun for the 
presence of NP in $B^0_s$--$\bar B^0_s$ mixing will be the detection 
of a non-vanishing value of $\phi_s^{\rm NP}$ through CP violation in 
$B_s\to J/\psi\phi$. 
This example is yet another demonstration that flavour physics is
not an optional extra, but an indispensable
ingredient in the pursuit of NP, also and in particular in the era of the LHC.

\section*{Acknowledgments}
It is a pleasure to thank R.~Fleischer for a very enjoyable collaboration on
the work presented here, and the organisers of the Moriond meetings
for the invitation. This work was supported in part by the EU networks
contract Nos.\ MRTN-CT-2006-035482, {\sc Flavianet}, and
MRTN-CT-2006-035505, {\sc Heptools}.

\end{document}